\documentclass[%
 reprint,
 amsmath,amssymb,
 aps,
 prl,
]{revtex4-1}

\usepackage{graphicx}
\usepackage{dcolumn}
\usepackage{bm}
\usepackage{epsfig,color,xspace,multirow,xr,bbold}
\usepackage[all]{xy}
\usepackage{setspace}
\usepackage{url}

\usepackage{hyperref}
\hypersetup{
   colorlinks,
   menucolor=blue,
   linkcolor=black,
   citecolor=blue,
   urlcolor=blue
}

\newcommand*\diff{\mathop{}\!\mathrm{d}}

\begin{document}


\title{Microscopic reweighting for non-equilibrium steady states dynamics}
 
\author{Marius Bause}
\email{bause@mpip-mainz.mpg.de}
\author{Timon Wittenstein}%
\author{Kurt Kremer}%
 \author{Tristan Bereau}%
\affiliation{%
Max Planck Institute for Polymer Research, 55128 Mainz, Germany
}%

\date{\today}

\begin{abstract}
  Computer simulations generate trajectories at a single, well-defined
  thermodynamic state point. Statistical reweighting offers the means
  to reweight static and dynamical properties to different equilibrium
  state points by means of analytic relations. We extend these ideas
  to non-equilibrium steady states by relying on a maximum path
  entropy formalism subject to physical constraints. Stochastic
  thermodynamics analytically relates the forward and backward
  probabilities of any pathway through the external non-conservative
  force, enabling reweighting both in and out of equilibrium.  We avoid
  the combinatorial explosion of microtrajectories by systematically
  constructing pathways through Markovian transitions.  We further
  identify a quantity that is invariant to dynamical reweighting,
  analogous to the density of states in equilibrium reweighting.
\end{abstract}

\maketitle

Many chemical and biological processes are influenced by external
driving forces and operate away from equilibrium---examples include
colloidal particles, biopolymers, enzymes, or molecular motors
\cite{SeifertRev}. Despite our current lack of a universal theory for
statistical mechanics off equilibrium~\cite{dougherty1994foundations},
computer simulations can complement experiments by providing
microscopic insight into these complex processes.  Unfortunately
current computational power often prevents molecular simulations from
reaching the experimentally-relevant time scales, or alternatively,
obliges them to operate at artificially-large driving
forces~\cite{perilla2015molecular}.  The latter motivates a formalism
to reweight dynamics across off-equilibrium conditions.

When dealing with systems in equilibrium, Ferrenberg and Swendsen
introduced a statistical-reweighting procedure to infer information
about a system when sampled at another state point~\cite{ferrenberg88,
ferrenberg89}. It requires microscopic information at fixed
thermodynamic conditions, e.g., temperature, collected by computer
simulations or experiments. A probability associated with each microstate
is reweighted according to physical
relationships linking the initial and final thermodynamic conditions.
Reweighting can be conducted arbitrarily far from the initial state, 
provided it is sufficiently sampled.

Equilibrium reweighting has led to a number of developments in the
field, from estimating accurate free energies~\cite{WHAM} to building
more robust Markov state models~\cite{RewDyn, dTRAM}.  In this Letter,
we generalize reweighting to dynamical processes by replacing
microstates with microtrajectories. The proposed methodology, valid
for non-equilibrium steady state (NESS) systems, employs a maximum
path entropy formalism while generalizing the standard detailed
balance relation.

Jaynes' maximum entropy approach offers a general variational
principle to understand macroscopic phenomena from microscopic
knowledge of a statistical system~\cite{jaynes}. This
information-theoretic method regards entropy as the measure of
uncertainty of the system.  Consider a coordinate $x$ of a system with
unknown probability distribution, $p(x)$.  We further define another
distribution, $q(x)$, used as a prior on $p(x)$.  The most likely
representation of $p(x)$ can be found by minimizing the cross-entropy
functional 
\begin{equation}
  \mathcal{C}[p(x)] = - \int \diff x \; p(x) 
    \ln  \left ( \frac{p(x)}{q(x)} \right ).
  \label{eq:KL} 
\end{equation}
This quantity was shown to fulfill the axioms for an uncertainty
measure \cite{SJ}.  Setting a uniform prior (i.e., $q(x) =
\text{const.}$) reduces to the well-known Shannon entropy and
minimizing the cross entropy in this case is equivalent to maximizing
the Shannon entropy. 

According to Jaynes, a system would maximize the number of microscopic
realizations compatible with a certain macroscopic state, linking the
two scales via constraints.  For instance, working in the canonical
ensemble will lead to a constraint on the average energy $\langle E
\rangle$
\begin{equation}
\begin{aligned}
\mathcal{C_{\text{equ}}} = - \int \diff x \; p(x) \ln \left ( \frac{p(x)}{q(x)} \right )  - \zeta \left ( \int \diff x \; p(x) - 1 \right ) \\
- \beta \left( \int \diff x \; p(x) E(x) - \langle E \rangle \right),
\end{aligned}
\label{eq:Caliber1}
\end{equation}
where $\zeta$ and $\beta$ are Lagrangian multipliers, controlling the
normalization of probabilities and the fixed average energy.
Minimization of $\mathcal{C_{\text{equ}}} $ with respect to $p(x)$
yields
\begin{equation}
 p(x) = \frac{q(x) }{\tilde Z(\beta)} \exp 
  \left( - \beta E(x) \right ), 
 \label{eq:FSrew}
\end{equation}
where the partition function $\tilde Z (\beta) = \int \diff x \, q(x)
\exp \left (- \beta E(x) \right ) $ becomes a normalization constant
and the Lagrange multiplier $\beta$ is identified with the inverse
temperature, $\beta^{-1} = k_{\rm B}T$.  This approach naturally lends
itself to reweighting: Given a reference distribution $q(x)$ sampled
at inverse temperature $\beta'$, microscopic information at a third
inverse temperature $\beta'' = \beta + \beta' $ is inferred through
the calculation of $p(x)$. This reweighting becomes exact under full
knowledge of the density of states function $\Omega(x) = q(x) \exp
\left (  \beta' E(x) \right )$.

The maximum entropy formalism has been generalized to the study of
dynamical systems by working with microtrajectories---an approach
called Maximum Caliber~\cite{jaynesCaliber}. It was shown to recover
known off-equilibrium relations~\cite{dixitMaxCal} and predict
dynamical pathways in NESSs correctly when supplied with appropriate
constraints \cite{agozzino2019dynamics}. Conceptually the approach
follows the same scheme as the previous derivation of equilibrium
reweighting: The most likely microtrajectories maximize the path
entropy function subject to physical constraints. 

The following discusses a rich and relevant subset of non-equilibrium
processes: non-equilibrium steady states.  NESS correspond to the
long-time limit under constant driving by an external
reservoir~\cite{zia2007probability}. As such, time symmetry is broken,
but the fluxes within the system are time independent, and so are the
distributions of microtrajectories.

Compared to equilibrium reweighting that focuses on the sampling of
microstates, dynamical reweighting of NESS considers
microtrajectories---collections of microstates---which become
computationally intractable for all but the smallest of systems. To
complicate things further, the length of time of a microtrajectory is
a priori unknown.  To resolve these issues, we map all trajectories to
a first-order Markov process.  This coarse-graining of
microtrajectories leaves us with the easier task of sampling
transition probabilities, and subsequently \emph{constructing}
microtrajectories out of the combination of individual
microtransitions. Such an approach facilitates the sampling of
microtrajectories.

Markov state models (MSM) discretize configurational space into
so-called microstates (i.e., collection of microscopic states) as well
as time in terms of steps of constant length $\tau$ (i.e., the
lagtime), thereby mapping  trajectories to a discrete-time Markov
Chain \cite{BookNoe}. All observed transitions are collected to infer
a transition probability matrix $p_{ij} (\tau)$, where $i$ and $j$
label microstates.  An appropriate space and time discretization helps
fulfill the Markovian assumption~\cite{prinz2011markov}. Markov state
models have proven powerful tools for reaching time scales that are
unattainable by brute-force computer
simulations~\cite{plattner2017complete}.

Utilizing the Markovian assumption, the microtrajectories of the
abovementioned cross-entropy functional (Eqn.~\ref{eq:KL}) reduces to
\begin{equation}
  \mathcal{C} = -\sum_{i,j} \pi_i p_{ij}\ln \frac{p_{ij}}{q_{ij}},
 \label{eq:Calibernew}
\end{equation}
where $\mathbf{\pi}$ corresponds to the stationary
distribution~\cite{derivationCal}. In the absence of constraints, the
minimum of the cross entropy is its prior $q_{ij}$. Previous work has
shown how to constrain the system according to microscopic and/or
macroscopic constraints: ($i$) the matching of simulation and
experimental data at equilibrium by enforcing detailed
balance~\cite{rudzinski2016communication, CCMM}; ($ii$) inferring
kinetic rates given variations in equilibrium
populations~\cite{MaxCalMutation}; or ($iii$) by using the stationary
distribution and macroscopic constraints corresponding to a NESS
experiment~\cite{dixit2015inferring}. Such macroscopic constraints are
typically process dependent and not always known. In the current
Letter, we propose to constrain dynamics to NESS by drawing
\emph{microscopic} balance constraints from stochastic thermodynamics
and enforce them on the cross entropy (Eqn.~\ref{eq:Calibernew}). We
consider several constraints:  The conservation of probability flow
through so-called global balance, $\sum_i \pi_j p_{ji} = \sum_i \pi_i
p_{ij}$, allowing for global fluxes in the system; Normalization
considerations imply $\sum_j p_{ij} = 1$ and $\sum_i \pi_{i} = 1$.
Since all transition probabilities in NESSs are time-independent the
existence of a steady-state distribution $\mathbf{\pi}$ is
guaranteed~\cite{evans2010probability,evans2016relaxation}.

Furthermore, we want to constrain the system according to microscopic 
reversibility~\cite{crooks2011thermodynamic,CrooksMicRev}
\begin{equation}
\frac{\mathcal{P}[\Gamma(+t)| f(+t)]}{ 
  \mathcal{P}[ \bar \Gamma(-t)| \bar f(-t) ]} = 
  \exp \left( -\beta Q| \Gamma(+t)| f(+t)] \right),
\label{eq:microrev} 
\end{equation}
where $\mathcal{P}[ \Gamma(+t)| f(+t)]$ denotes the probability of observing
the time-forward trajectory $x(+t)$ under the external driving force
$f(+t)$, $\mathcal{P}[\bar \Gamma(-t)| \bar f(-t) ]$ points at the 
time-reversed trajectory, while $Q[\Gamma(+t)| f(+t)]$ refers to the amount
of heat exchanged between the system and the reservoir along a given
trajectory and acting forces. Critically, this links the probability
of a forward trajectory with its time-reversed counterpart. In case of
equilibrium dynamics, the relation becomes path independent and
simplifies to detailed balance~\cite{zhang2012stochastic}.
For a more general expression, we integrate Eqn.~\ref{eq:microrev} 
over the complete set of initial states $i$, target
states $j$, as well as the set of all trajectories connecting them, to
obtain the coarser expression (see derivation in S1)
\begin{equation}
\langle \Delta S_{ij} \rangle = \ln \frac{p_{ij}}{p_{ji}},
\label{eq:Ssim}
\end{equation}
where $\Delta S_{ij}$ is called local entropy production, describing 
the amount of work an external reservoir has to
perform on the system to transition between two states. 
This quantity naturally
generalizes detailed balance~\cite{maes2007and} and will be used as a microscopic
constraint on the Caliber.

Having defined all constraints, the Caliber functional becomes
\begin{widetext}
\begin{equation}
       \mathcal{C_{\text{dyn}}} = \underbrace{-\sum_{i,j} \pi_i p_{ij}\ln 
\frac{p_{ij}}{q_{ij}} }_{\text{Caliber}}
      +  \underbrace{ \sum_i \mu_i \left( \sum_j p_{ij} - 1 \right) + \zeta 
\sum_i ( \pi_i -1) }_{\text{Normalization}} 
      + \underbrace{ \sum_j \nu_j \left(\sum_i \pi_i p_{ij} - \pi_j \right) 
}_{\text{Global Balance}}
      + \underbrace{ \sum_{ij}^{i<j} \pi_i  \alpha_{ij} \left( \ln \left( 
\frac{p_{ij}}{p_{ji}} \right) - \Delta S_{ij} \right)}_{\text{Local Balance}},
    \label{eq:Caliber2}
\end{equation}
\end{widetext}
where $\mu_i$, $\zeta$, $\nu_j$ and $\alpha_{ij}$ are Lagrange
multipliers.  Eqn.~\ref{eq:Caliber2} modifies the
equilibrium-reweighting Caliber (Eqn.~\ref{eq:Caliber1}) in several
ways: ($i$) The entropy expression is discretized and replaced by the
path entropy; ($ii$) transition probabilities are normalized; ($iii$)
a global-balance condition ensures a steady state; and ($iv$) local
entropy production is introduced as a NESS extension to detailed
balance. The solution is obtained by minimizing with respect to the
set of transition probabilities and the stationary distribution.
Assuming that $\Delta S_{ij}$ is small (see derivation in S2), we
obtain
\begin{equation}
\begin{aligned}
p_{ij} = q_{ij} \exp \left ( \frac{c_i + c_j}{2}   
+ \frac{\Delta S_{ij} - \Delta S^q_{ij}}{2} + \zeta \right ),
\end{aligned}
\label{eq:finalP} 
\end{equation}
which only depends on $\Delta S^q_{ij} = \ln \left( q_{ij} / q_{ji}
\right)$ from the reference simulation, $\Delta S_{ij}$,  and the
unknown constants $c_i$.  We note that $\Delta S_{ij}$ corresponds to
the local entropy production in the target state. Analogous to
histogram reweighting, the unknown coefficients $c_i$ can be found via
nonlinear relationships~\cite{bereau2009optimized}
\begin{equation}
1 = \sum_j p_{ij}  = \sum_j \sqrt{q_{ij}
q_{ji} }
\exp \left ( \frac{c_i + c_j + 2 \zeta }{2}   + \frac{\Delta S_{ij}}{2}
\right ).
\end{equation}
The set of equations is convex (see S4) and is solved by self
iteration from a randomly-selected starting point until convergence.

An alternative formalism by means of the Girsanov theorem was
introduced, which relies on single-trajectory probability reweighting
and the Boltzmann distribution to estimate the change between
equilibrium state points~\cite{donati2017girsanov}. In another work,
equilibrium transition rates are reweighted by a Maximum Caliber
formalism enforcing the Boltzmann distribution~\cite{MaxCalMutation}.
In contrast, the present method reweights MSMs in NESS without prior
knowledge of the steady-state distribution, but rather through the
entropy production. 

\begin{figure}[htbp]
  \centering
  \includegraphics[width = 0.85\linewidth]{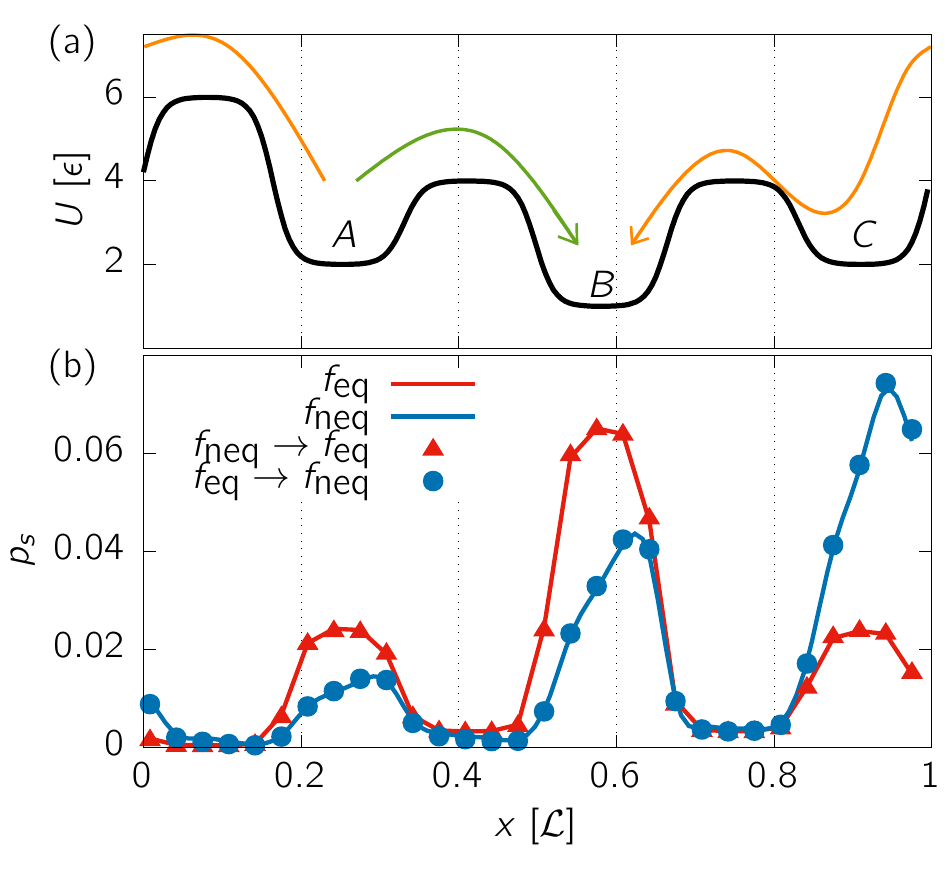}
  \caption{(a) Periodic potential energy surface experienced by the
    particle. The orange and green line show possible paths from
    metastable state A to B. (b) Stationary distributions for the
    equilibrium reference system, $f_{\rm eq} = 0$ (red line), and
    under the influence of a non-conservative driving force, $f_{\rm
    neq} = 9~\epsilon / \mathcal{L}$ (blue line). The points show the
    probability distributions obtained by sampling in one of the two
    states and reweighting into the other, $f_{\rm neq} \rightarrow
    f_{\rm eq}$ and $f_{\rm eq}
    \rightarrow f_{\rm neq}$.}
  \label{fig:potential}
\end{figure}

\emph{Application.} The reweighting procedure is tested on a single
particle driven by a non-conservative force $f$ along a
periodic one-dimensional potential $U(x)$ (see
Fig.~\ref{fig:potential}). The non-conservative force may emerge from
magnetic fields, mechanical flows, or mechanical dragging. An
analogous setup was experimentally studied, using silica spheres on a
tilted surface \cite{ma2017colloidal}. The overdamped equation of
motion for the particle is given by
\begin{equation}
0 = -\frac{\partial U( x)}{\partial x} - \gamma \dot x+ \sqrt{2\gamma k_{\rm B} T} 
R(t) + f,
\end{equation}
where $T$ is the temperature of a canonical reservoir coupled to the
system by friction constant $\gamma$.  $R(t)$ is a $\delta$-correlated
Gaussian process with mean $0$. Both the temperature and the potential
energy are fixed, while the reweighting is performed over different
ranges of non-conservative forces. We report results in reduced units,
where the box size is set to $\mathcal{L}$, the mass of the particle
is set to $\mathcal{M}$, and energy is measured in $\epsilon$. The
temperature is chosen to be $T = 1~ \epsilon / k_{\rm B}$, the energy
barriers shown in Fig.~\ref{fig:potential} are $2-4~ k_{\rm B}T$ .
Following our Markovian approximation, the model is separated in 60
microstates of equal size and a lagtime $\tau = 8 \cdot 10^{-4}
\mathcal{T}$, where $\mathcal{T} = \mathcal{L}
\sqrt{\mathcal{M}/{\epsilon}}$ is the unit of time.  The integration
time step was set to $\delta t = 10^{-5} ~\mathcal{T}$.  The
non-conservative force is varied between 0 and $9~\epsilon /
\mathcal{L}$.

For the given model, we want to derive an analytic
expression for the local entropy productions $\Delta S_{ij}$, required
by the reweighting procedure (Eqn.~\ref{eq:finalP}). Given an
underlying force field $\mathbf{F} = - \nabla U (x) + 
\mathbf{f}$, the local entropy production of a
single continuous trajectory $\mathbf{\Gamma}(t)$ is given by 
\begin{equation}
  \label{eq:analytic}
  \Delta S [\Gamma (t)] = \int \diff t 
  \frac{ \mathbf{F} \cdot \mathbf{ \dot \Gamma}}{T},
\end{equation}
where the quantity is integrated over time, $\mathbf{ \dot \Gamma}$ is
the velocity and $T$ is the temperature~\cite{seifert2005entropy}.
Assuming a constant non-conservative force $\mathbf{f}$ and making
use of a numerically discretized trajectory, $\mathbf{\Gamma}(t) \approx
\{\mathbf{x}_k\}$, we approximate $\Delta S$ between starting and target points
$x_i$ and $x_j$, respectively,
\begin{equation}
 \Delta S_{ij} (\{\mathbf{x}_k\}) \approx \frac{ U(x_i) - U(x_j) + 
 \sum_k (\mathbf{x}_{k+1} - \mathbf{x}_{k}) \cdot \mathbf{f}   }{ k_{\rm B}T }.
 \label{eq:Sprodth1}
\end{equation}
Because the entropy production of forward and backward steps directly
cancel, the quantity is unaffected by path variations in one
dimension.  Still, the periodic boundary conditions permit two
different results between $i$ and $j$, as indicated in
Fig.~\ref{fig:potential}: the shorter and longer paths (green and
orange, respectively), such that Eqn.~\ref{eq:Sprodth1} has two
solutions. By choosing the lagtime of the MSM reasonably short, we
effectively scale down the longer paths to a negligible weight. As
such, our expression for the local entropy production becomes
\begin{equation}
  \Delta S_{ij} \approx \frac{ U(x_i) - U(x_j) + (x_j -x_i) f  }{ 
 k_{\rm B}T }.
 \label{eq:Sprodth}
\end{equation}
We find excellent agreement between this expression and
Eqn.~\ref{eq:Ssim} when directly sampled from an MSM: a weighted
average error of 1\%, which does not affect the quality of the
reweighting upon insertion in Eqn.~\ref{eq:finalP}. A detailed
comparison is illustrated in S3. 

To assess our reweighting procedure, we monitor both static and
kinetic properties: ($i$) the stationary distribution of the
particle position and ($ii$) the first-passage-time distributions
between metastable states.  The metastable states are labeled $A$,
$B$, and $C$ (Fig.~\ref{fig:potential}). 

Figure \ref{fig:potential} shows the stationary distributions of the
particle position both in equilibrium and under the influence of a
driving force.  We reweight the simulation data from equilibrium to
the NESS and vice versa, demonstrating that the correct static
distributions are recovered when reweighting both in and out of
equilibrium---a result that holds for any pair of state points as
described further below.  

\begin{figure}
  \centering
  \includegraphics[width = 0.85 \linewidth]{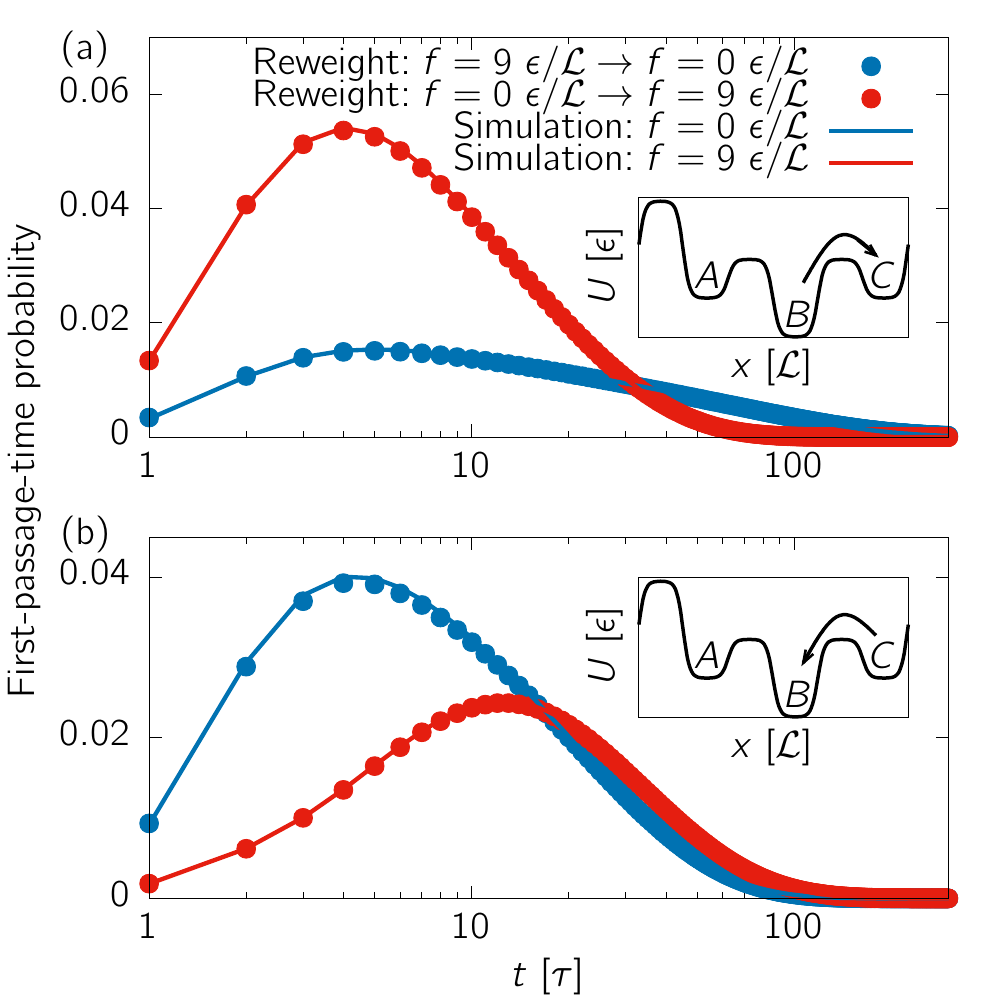}
  \caption{First Passage Time probabilities between metastable states
    B and C, expressed as a function of time (in units of the lagtime
    $\tau$).  The lines show the simulation data at $f=0$ and 
    $f=9~\epsilon / \mathcal{L}$. The points show the results under
    reweighting from each other for 
    processes (a) $B \rightarrow C$ and (b) $C \rightarrow B$. }
  \label{fig:FPTD}
\end{figure}

Turning to dynamical properties, Fig.~\ref{fig:FPTD} reports the
first-passage-time distributions between two metastable states in
equilibrium and under a strong driving force ($f=9~\epsilon /
\mathcal{L})$. The change in the broadness of the distributions
(Fig.~\ref{fig:FPTD}a) and the shift in the peak position
(Fig.~\ref{fig:FPTD}b) suggest that the set of dominant trajectories
change significantly under driving. This example shows that the
external driving force changes both the timescale and corresponding
processes of a transition. Here again, the reweighting procedure
recovers the first-passage-time distributions in either directions:
from equilibrium to NESS and vice versa.

\begin{figure}
 \includegraphics[width = 0.85 \linewidth]{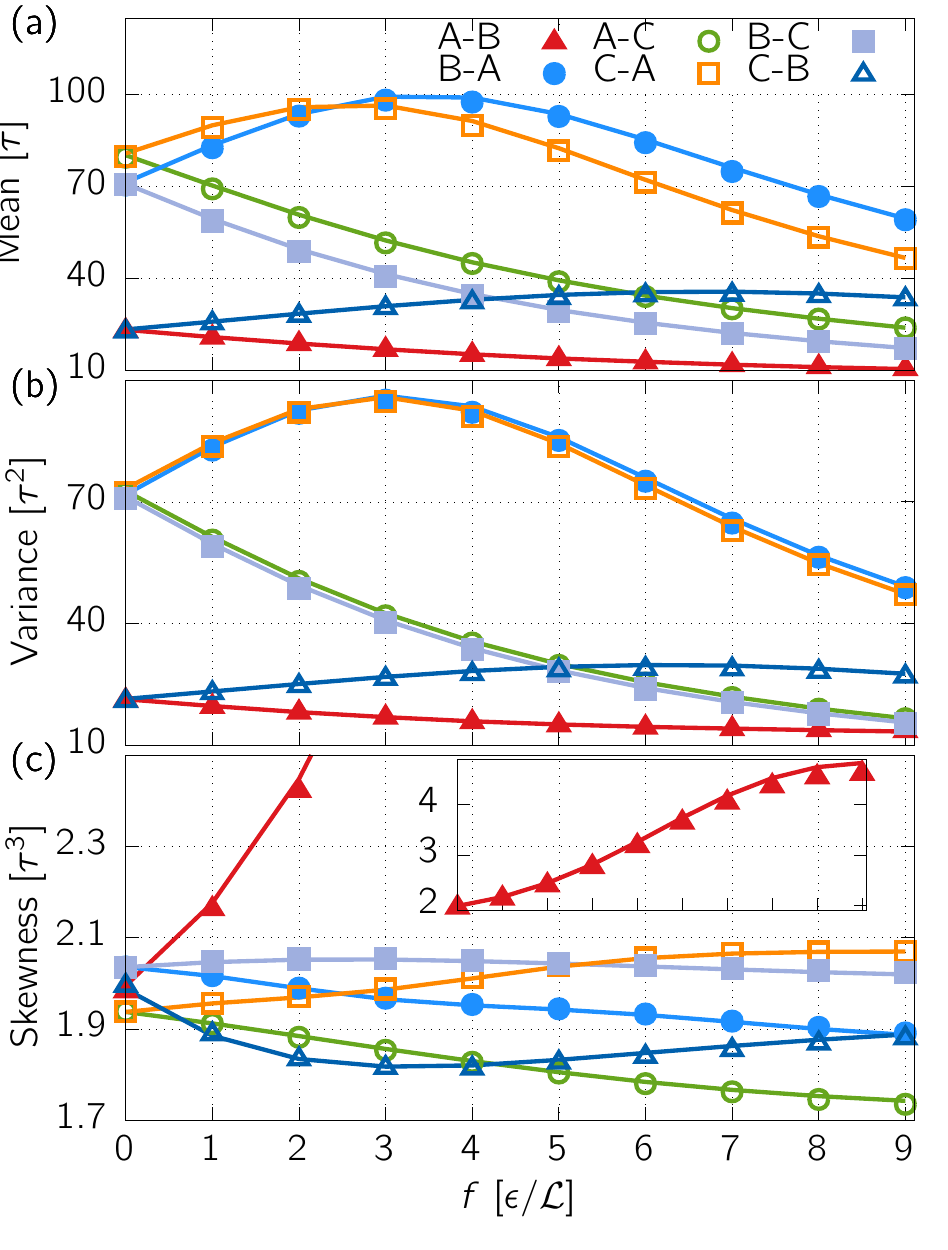}
 \caption{Moments of the first-passage-time distribution between three
   metastable states under external driving: (a) mean, (b) variance,
   and (c) skewness.  The metastable states are defined in figure
   \ref{fig:potential}.  The points correspond to reference
   simulations at various states, while the lines show the quantity
   under continuous reweighting, always choosing the equilibrium
   system ($f=0$) as reference.}
  \label{fig:rew}
\end{figure}

This analysis is extended to other state points by comparing the mean,
variance, and skewness of the first-passage-time distributions in
Fig.~\ref{fig:rew}.  The equilibrium system is chosen as a reference
and is continuously reweighted to off-equilibrium driven systems, even
though any other reference state point could be selected. Additionally, 
the reweighting procedure 
is applied to equilibrium systems under variation of potential (see S6 for results). The
reference potential energy surface (Fig.~\ref{fig:potential}) is such that
several pairs of processes show the same dynamics at equilibrium: The
transition $A \rightarrow B$ and $C \rightarrow  B$, but also
$B\rightarrow A$ and $B\rightarrow C$ as well as $A\rightarrow C$ and
$C\rightarrow A$.

Upon driving the system off-equilibrium, these symmetries break---a
phenomenon captured by the reweighting procedure. Increasing $f$
strongly affects the mean-first-passage times (Fig.~\ref{fig:rew}),
thereby altering the nature of the slowest processes.  We first
analyze the metastable transitions situated along the direction of
$f$. While the processes $A\rightarrow B$ and
$B\rightarrow C$ speed up with increasing driving, $C\rightarrow A$
shows a non-monotonic behavior: it first slows down up to a driving
force $f \approx 4~\epsilon / \mathcal{L}$ before speeding up. The
variance reveals that a broader selection of trajectories becomes
dominant before the threshold. Turning to metastable transitions that
oppose to the driving force, only the process $C\rightarrow B$ slows
down with driving, while $A\rightarrow C$ constantly speeds up,
despite unfavorable driving. Process $B\rightarrow A$ shows
non-monotonic behavior, similar to $C\rightarrow A$. This
counterintuitive behavior can be explained by the growing number of
long transitions from $B$ to $A$ via $C$. 

Reweighting implies the existence of an invariant quantity,
irrespective of the state point or driving force. Here we can isolate
the following invariant $I_{ij} = \sqrt{q_{ij} q_{ji} } \exp  \left
(\frac{c_i + c_j}{4} + \zeta \right )$ (see S5 for derivation).  We
can thereby rewrite the abovementioned solution of the Caliber
(Eqn.~\ref{eq:finalP})) as
\begin{equation}
  p_{ij} = \frac{ I_{ij} } {Z_{ij}}  
  \exp{ \left( \frac{ \Delta S_{ij} }{2} \right )},
  \label{eq:invariant}
  \end{equation}
using the normalization $Z_{ij} = \exp{ \left ( \frac{   -c_i -
c_j}{4} \right )}$. Note that $Z_{ij}$ depends on both $I_{ij}$ and
$\Delta S_{ij}$ via the relation $ 1 = \sum_j \frac{ I_{ij} } {Z_{ij}}
\exp{\left(\frac{ \Delta S_{ij} }{2} \right )}$ and accounts for the
interconnection of the states. We draw similarities with equilibrium
reweighting in Eqn.~\ref{eq:FSrew}: ($i$) The probability is
proportional to the product of an invariant and an exponential
function (the density of states and the Boltzmann factor in
equilibrium); ($ii$) The partition function depends on the control
variable ($T$ or $\Delta S_{ij}$); and ($iii$) The reweighting only
depends on \emph{relative} quantities, only requiring knowledge of
temperature difference or changes in the local entropy production.
Both procedures show striking similarities in their derivation,
functional form, and properties. 

The present reweighting method is a generalization of existing
Likelihood and Maximum Caliber methods that have been applied to
systems in and out of equilibrium with varying microscopic and
macroscopic constraints. The microscopic expression for the local
entropy production acts as a local constraint that generalizes
detailed balance for NESS. We show that this choice governs static and
dynamic properties of a NESS and enables us to reproduce these
properties over a wide range of driving. The analytic expression for
the relative entropy production allows us to continuously tune the
external driving force and quantitatively reweight the stationary
distribution and kinetic properties. 


The Maximum Caliber formalism in combination with local entropy
productions offer an analytic relation between NESSs. Dynamical data
of a system can be gathered in a driving-invariant quantity and
detailed kinetic information at any thermodynamic state point can be
recovered. This idea allows to populate rare transition
paths~\cite{bolhuis2002transition}: A driving force may push the
system to discover new paths, the reweighting procedure recovers
detailed dynamical information of the sampled path at any another
thermodynamic state point. In case equilibrium dynamics are of
interest, the entropy productions only depend on the free energy. Low
weight trajectories can thus be calculated with high accuracy and no
further information. By tuning the relative local entropy productions
of the system, the reweighting allows to study dynamical properties
and pathways in NESS without further simulation.

\section*{Acknowledgments}

We thank Joseph F. Rudzinski, Dominik Spiller and Johannes Zierenberg for insightful discussions.  This 
work was supported in part by the Emmy Noether program of the Deutsche
Forschungsgemeinschaft (DFG) to TB, the Graduate School of Excellence
Materials Science in Mainz (MAINZ) to MB, and the European Research
Council under the European Union's Seventh Framework Programme
(MOLPROCOMP) to KK.

\bibliographystyle{apsrev4-1}
\bibliography{mybib.bib}

\begin{thebibliography}{33}%
\makeatletter
\providecommand \@ifxundefined [1]{%
 \@ifx{#1\undefined}
}%
\providecommand \@ifnum [1]{%
 \ifnum #1\expandafter \@firstoftwo
 \else \expandafter \@secondoftwo
 \fi
}%
\providecommand \@ifx [1]{%
 \ifx #1\expandafter \@firstoftwo
 \else \expandafter \@secondoftwo
 \fi
}%
\providecommand \natexlab [1]{#1}%
\providecommand \enquote  [1]{``#1''}%
\providecommand \bibnamefont  [1]{#1}%
\providecommand \bibfnamefont [1]{#1}%
\providecommand \citenamefont [1]{#1}%
\providecommand \href@noop [0]{\@secondoftwo}%
\providecommand \href [0]{\begingroup \@sanitize@url \@href}%
\providecommand \@href[1]{\@@startlink{#1}\@@href}%
\providecommand \@@href[1]{\endgroup#1\@@endlink}%
\providecommand \@sanitize@url [0]{\catcode `\\12\catcode `\$12\catcode
  `\&12\catcode `\#12\catcode `\^12\catcode `\_12\catcode `\%12\relax}%
\providecommand \@@startlink[1]{}%
\providecommand \@@endlink[0]{}%
\providecommand \url  [0]{\begingroup\@sanitize@url \@url }%
\providecommand \@url [1]{\endgroup\@href {#1}{\urlprefix }}%
\providecommand \urlprefix  [0]{URL }%
\providecommand \Eprint [0]{\href }%
\providecommand \doibase [0]{http://dx.doi.org/}%
\providecommand \selectlanguage [0]{\@gobble}%
\providecommand \bibinfo  [0]{\@secondoftwo}%
\providecommand \bibfield  [0]{\@secondoftwo}%
\providecommand \translation [1]{[#1]}%
\providecommand \BibitemOpen [0]{}%
\providecommand \bibitemStop [0]{}%
\providecommand \bibitemNoStop [0]{.\EOS\space}%
\providecommand \EOS [0]{\spacefactor3000\relax}%
\providecommand \BibitemShut  [1]{\csname bibitem#1\endcsname}%
\let\auto@bib@innerbib\@empty
\bibitem [{\citenamefont {Seifert}(2008)}]{SeifertRev}%
  \BibitemOpen
  \bibfield  {author} {\bibinfo {author} {\bibfnamefont {U.}~\bibnamefont
  {Seifert}},\ }\href@noop {} {\bibfield  {journal} {\bibinfo  {journal} {The
  European Physical Journal B}\ }\textbf {\bibinfo {volume} {64}},\ \bibinfo
  {pages} {423} (\bibinfo {year} {2008})}\BibitemShut {NoStop}%
\bibitem [{\citenamefont {Dougherty}(1994)}]{dougherty1994foundations}%
  \BibitemOpen
  \bibfield  {author} {\bibinfo {author} {\bibfnamefont {J.~P.}\ \bibnamefont
  {Dougherty}},\ }\href@noop {} {\bibfield  {journal} {\bibinfo  {journal}
  {Philosophical Transactions of the Royal Society of London. Series A:
  Physical and Engineering Sciences}\ }\textbf {\bibinfo {volume} {346}},\
  \bibinfo {pages} {259} (\bibinfo {year} {1994})}\BibitemShut {NoStop}%
\bibitem [{\citenamefont {Perilla}\ \emph {et~al.}(2015)\citenamefont
  {Perilla}, \citenamefont {Goh}, \citenamefont {Cassidy}, \citenamefont {Liu},
  \citenamefont {Bernardi}, \citenamefont {Rudack}, \citenamefont {Yu},
  \citenamefont {Wu},\ and\ \citenamefont {Schulten}}]{perilla2015molecular}%
  \BibitemOpen
  \bibfield  {author} {\bibinfo {author} {\bibfnamefont {J.~R.}\ \bibnamefont
  {Perilla}}, \bibinfo {author} {\bibfnamefont {B.~C.}\ \bibnamefont {Goh}},
  \bibinfo {author} {\bibfnamefont {C.~K.}\ \bibnamefont {Cassidy}}, \bibinfo
  {author} {\bibfnamefont {B.}~\bibnamefont {Liu}}, \bibinfo {author}
  {\bibfnamefont {R.~C.}\ \bibnamefont {Bernardi}}, \bibinfo {author}
  {\bibfnamefont {T.}~\bibnamefont {Rudack}}, \bibinfo {author} {\bibfnamefont
  {H.}~\bibnamefont {Yu}}, \bibinfo {author} {\bibfnamefont {Z.}~\bibnamefont
  {Wu}}, \ and\ \bibinfo {author} {\bibfnamefont {K.}~\bibnamefont
  {Schulten}},\ }\href@noop {} {\bibfield  {journal} {\bibinfo  {journal}
  {Current opinion in structural biology}\ }\textbf {\bibinfo {volume} {31}},\
  \bibinfo {pages} {64} (\bibinfo {year} {2015})}\BibitemShut {NoStop}%
\bibitem [{\citenamefont {Ferrenberg}\ and\ \citenamefont
  {Swendsen}(1988)}]{ferrenberg88}%
  \BibitemOpen
  \bibfield  {author} {\bibinfo {author} {\bibfnamefont {A.~M.}\ \bibnamefont
  {Ferrenberg}}\ and\ \bibinfo {author} {\bibfnamefont {R.~H.}\ \bibnamefont
  {Swendsen}},\ }\href@noop {} {\bibfield  {journal} {\bibinfo  {journal}
  {Physical review letters}\ }\textbf {\bibinfo {volume} {61}},\ \bibinfo
  {pages} {2635} (\bibinfo {year} {1988})}\BibitemShut {NoStop}%
\bibitem [{\citenamefont {Ferrenberg}\ and\ \citenamefont
  {Swendsen}(1989)}]{ferrenberg89}%
  \BibitemOpen
  \bibfield  {author} {\bibinfo {author} {\bibfnamefont {A.~M.}\ \bibnamefont
  {Ferrenberg}}\ and\ \bibinfo {author} {\bibfnamefont {R.~H.}\ \bibnamefont
  {Swendsen}},\ }\href@noop {} {\bibfield  {journal} {\bibinfo  {journal}
  {Computers in Physics}\ }\textbf {\bibinfo {volume} {3}},\ \bibinfo {pages}
  {101} (\bibinfo {year} {1989})}\BibitemShut {NoStop}%
\bibitem [{\citenamefont {Kumar}\ \emph {et~al.}(1992)\citenamefont {Kumar},
  \citenamefont {Rosenberg}, \citenamefont {Bouzida}, \citenamefont
  {Swendsen},\ and\ \citenamefont {Kollman}}]{WHAM}%
  \BibitemOpen
  \bibfield  {author} {\bibinfo {author} {\bibfnamefont {S.}~\bibnamefont
  {Kumar}}, \bibinfo {author} {\bibfnamefont {J.~M.}\ \bibnamefont
  {Rosenberg}}, \bibinfo {author} {\bibfnamefont {D.}~\bibnamefont {Bouzida}},
  \bibinfo {author} {\bibfnamefont {R.~H.}\ \bibnamefont {Swendsen}}, \ and\
  \bibinfo {author} {\bibfnamefont {P.~A.}\ \bibnamefont {Kollman}},\
  }\href@noop {} {\bibfield  {journal} {\bibinfo  {journal} {Journal of
  computational chemistry}\ }\textbf {\bibinfo {volume} {13}},\ \bibinfo
  {pages} {1011} (\bibinfo {year} {1992})}\BibitemShut {NoStop}%
\bibitem [{\citenamefont {Chodera}\ \emph {et~al.}(2011)\citenamefont
  {Chodera}, \citenamefont {Swope}, \citenamefont {No{\'e}}, \citenamefont
  {Prinz}, \citenamefont {Shirts},\ and\ \citenamefont {Pande}}]{RewDyn}%
  \BibitemOpen
  \bibfield  {author} {\bibinfo {author} {\bibfnamefont {J.~D.}\ \bibnamefont
  {Chodera}}, \bibinfo {author} {\bibfnamefont {W.~C.}\ \bibnamefont {Swope}},
  \bibinfo {author} {\bibfnamefont {F.}~\bibnamefont {No{\'e}}}, \bibinfo
  {author} {\bibfnamefont {J.-H.}\ \bibnamefont {Prinz}}, \bibinfo {author}
  {\bibfnamefont {M.~R.}\ \bibnamefont {Shirts}}, \ and\ \bibinfo {author}
  {\bibfnamefont {V.~S.}\ \bibnamefont {Pande}},\ }\href@noop {} {\bibfield
  {journal} {\bibinfo  {journal} {The Journal of chemical physics}\ }\textbf
  {\bibinfo {volume} {134}},\ \bibinfo {pages} {06B612} (\bibinfo {year}
  {2011})}\BibitemShut {NoStop}%
\bibitem [{\citenamefont {Wu}\ \emph {et~al.}(2014)\citenamefont {Wu},
  \citenamefont {Mey}, \citenamefont {Rosta},\ and\ \citenamefont
  {No{\'e}}}]{dTRAM}%
  \BibitemOpen
  \bibfield  {author} {\bibinfo {author} {\bibfnamefont {H.}~\bibnamefont
  {Wu}}, \bibinfo {author} {\bibfnamefont {A.~S.}\ \bibnamefont {Mey}},
  \bibinfo {author} {\bibfnamefont {E.}~\bibnamefont {Rosta}}, \ and\ \bibinfo
  {author} {\bibfnamefont {F.}~\bibnamefont {No{\'e}}},\ }\href@noop {}
  {\bibfield  {journal} {\bibinfo  {journal} {The Journal of chemical physics}\
  }\textbf {\bibinfo {volume} {141}},\ \bibinfo {pages} {12B629\_1} (\bibinfo
  {year} {2014})}\BibitemShut {NoStop}%
\bibitem [{\citenamefont {Jaynes}(1957)}]{jaynes}%
  \BibitemOpen
  \bibfield  {author} {\bibinfo {author} {\bibfnamefont {E.~T.}\ \bibnamefont
  {Jaynes}},\ }\href@noop {} {\bibfield  {journal} {\bibinfo  {journal}
  {Physical review}\ }\textbf {\bibinfo {volume} {106}},\ \bibinfo {pages}
  {620} (\bibinfo {year} {1957})}\BibitemShut {NoStop}%
\bibitem [{\citenamefont {Shore}\ and\ \citenamefont {Johnson}(1980)}]{SJ}%
  \BibitemOpen
  \bibfield  {author} {\bibinfo {author} {\bibfnamefont {J.}~\bibnamefont
  {Shore}}\ and\ \bibinfo {author} {\bibfnamefont {R.}~\bibnamefont
  {Johnson}},\ }\href@noop {} {\bibfield  {journal} {\bibinfo  {journal} {IEEE
  Transactions on information theory}\ }\textbf {\bibinfo {volume} {26}},\
  \bibinfo {pages} {26} (\bibinfo {year} {1980})}\BibitemShut {NoStop}%
\bibitem [{\citenamefont {Jaynes}(1980)}]{jaynesCaliber}%
  \BibitemOpen
  \bibfield  {author} {\bibinfo {author} {\bibfnamefont {E.~T.}\ \bibnamefont
  {Jaynes}},\ }\href@noop {} {\bibfield  {journal} {\bibinfo  {journal} {Annual
  Review of Physical Chemistry}\ }\textbf {\bibinfo {volume} {31}},\ \bibinfo
  {pages} {579} (\bibinfo {year} {1980})}\BibitemShut {NoStop}%
\bibitem [{\citenamefont {Dixit}\ \emph {et~al.}(2018)\citenamefont {Dixit},
  \citenamefont {Wagoner}, \citenamefont {Weistuch}, \citenamefont
  {Press{\'e}}, \citenamefont {Ghosh},\ and\ \citenamefont
  {Dill}}]{dixitMaxCal}%
  \BibitemOpen
  \bibfield  {author} {\bibinfo {author} {\bibfnamefont {P.~D.}\ \bibnamefont
  {Dixit}}, \bibinfo {author} {\bibfnamefont {J.}~\bibnamefont {Wagoner}},
  \bibinfo {author} {\bibfnamefont {C.}~\bibnamefont {Weistuch}}, \bibinfo
  {author} {\bibfnamefont {S.}~\bibnamefont {Press{\'e}}}, \bibinfo {author}
  {\bibfnamefont {K.}~\bibnamefont {Ghosh}}, \ and\ \bibinfo {author}
  {\bibfnamefont {K.~A.}\ \bibnamefont {Dill}},\ }\href@noop {} {\bibfield
  {journal} {\bibinfo  {journal} {The Journal of chemical physics}\ }\textbf
  {\bibinfo {volume} {148}},\ \bibinfo {pages} {010901} (\bibinfo {year}
  {2018})}\BibitemShut {NoStop}%
\bibitem [{\citenamefont {Agozzino}\ and\ \citenamefont
  {Dill}(2019)}]{agozzino2019dynamics}%
  \BibitemOpen
  \bibfield  {author} {\bibinfo {author} {\bibfnamefont {L.}~\bibnamefont
  {Agozzino}}\ and\ \bibinfo {author} {\bibfnamefont {K.~A.}\ \bibnamefont
  {Dill}},\ }\href@noop {} {\bibfield  {journal} {\bibinfo  {journal} {arXiv
  preprint arXiv:1904.11426}\ } (\bibinfo {year} {2019})}\BibitemShut {NoStop}%
\bibitem [{\citenamefont {Zia}\ and\ \citenamefont
  {Schmittmann}(2007)}]{zia2007probability}%
  \BibitemOpen
  \bibfield  {author} {\bibinfo {author} {\bibfnamefont {R.}~\bibnamefont
  {Zia}}\ and\ \bibinfo {author} {\bibfnamefont {B.}~\bibnamefont
  {Schmittmann}},\ }\href@noop {} {\bibfield  {journal} {\bibinfo  {journal}
  {Journal of Statistical Mechanics: Theory and Experiment}\ }\textbf {\bibinfo
  {volume} {2007}},\ \bibinfo {pages} {P07012} (\bibinfo {year}
  {2007})}\BibitemShut {NoStop}%
\bibitem [{\citenamefont {Bowman}\ \emph {et~al.}(2013)\citenamefont {Bowman},
  \citenamefont {Pande},\ and\ \citenamefont {No{\'e}}}]{BookNoe}%
  \BibitemOpen
  \bibfield  {author} {\bibinfo {author} {\bibfnamefont {G.~R.}\ \bibnamefont
  {Bowman}}, \bibinfo {author} {\bibfnamefont {V.~S.}\ \bibnamefont {Pande}}, \
  and\ \bibinfo {author} {\bibfnamefont {F.}~\bibnamefont {No{\'e}}},\
  }\href@noop {} {\emph {\bibinfo {title} {An introduction to Markov state
  models and their application to long timescale molecular simulation}}},\
  Vol.\ \bibinfo {volume} {797}\ (\bibinfo  {publisher} {Springer Science \&
  Business Media},\ \bibinfo {year} {2013})\BibitemShut {NoStop}%
\bibitem [{\citenamefont {Prinz}\ \emph {et~al.}(2011)\citenamefont {Prinz},
  \citenamefont {Wu}, \citenamefont {Sarich}, \citenamefont {Keller},
  \citenamefont {Senne}, \citenamefont {Held}, \citenamefont {Chodera},
  \citenamefont {Sch{\"u}tte},\ and\ \citenamefont
  {No{\'e}}}]{prinz2011markov}%
  \BibitemOpen
  \bibfield  {author} {\bibinfo {author} {\bibfnamefont {J.-H.}\ \bibnamefont
  {Prinz}}, \bibinfo {author} {\bibfnamefont {H.}~\bibnamefont {Wu}}, \bibinfo
  {author} {\bibfnamefont {M.}~\bibnamefont {Sarich}}, \bibinfo {author}
  {\bibfnamefont {B.}~\bibnamefont {Keller}}, \bibinfo {author} {\bibfnamefont
  {M.}~\bibnamefont {Senne}}, \bibinfo {author} {\bibfnamefont
  {M.}~\bibnamefont {Held}}, \bibinfo {author} {\bibfnamefont {J.~D.}\
  \bibnamefont {Chodera}}, \bibinfo {author} {\bibfnamefont {C.}~\bibnamefont
  {Sch{\"u}tte}}, \ and\ \bibinfo {author} {\bibfnamefont {F.}~\bibnamefont
  {No{\'e}}},\ }\href@noop {} {\bibfield  {journal} {\bibinfo  {journal} {The
  Journal of chemical physics}\ }\textbf {\bibinfo {volume} {134}},\ \bibinfo
  {pages} {174105} (\bibinfo {year} {2011})}\BibitemShut {NoStop}%
\bibitem [{\citenamefont {Plattner}\ \emph {et~al.}(2017)\citenamefont
  {Plattner}, \citenamefont {Doerr}, \citenamefont {De~Fabritiis},\ and\
  \citenamefont {No{\'e}}}]{plattner2017complete}%
  \BibitemOpen
  \bibfield  {author} {\bibinfo {author} {\bibfnamefont {N.}~\bibnamefont
  {Plattner}}, \bibinfo {author} {\bibfnamefont {S.}~\bibnamefont {Doerr}},
  \bibinfo {author} {\bibfnamefont {G.}~\bibnamefont {De~Fabritiis}}, \ and\
  \bibinfo {author} {\bibfnamefont {F.}~\bibnamefont {No{\'e}}},\ }\href@noop
  {} {\bibfield  {journal} {\bibinfo  {journal} {Nature chemistry}\ }\textbf
  {\bibinfo {volume} {9}},\ \bibinfo {pages} {1005} (\bibinfo {year}
  {2017})}\BibitemShut {NoStop}%
\bibitem [{\citenamefont {Lee}\ and\ \citenamefont
  {Press{\'e}}(2012)}]{derivationCal}%
  \BibitemOpen
  \bibfield  {author} {\bibinfo {author} {\bibfnamefont {J.}~\bibnamefont
  {Lee}}\ and\ \bibinfo {author} {\bibfnamefont {S.}~\bibnamefont
  {Press{\'e}}},\ }\href@noop {} {\bibfield  {journal} {\bibinfo  {journal}
  {The Journal of chemical physics}\ }\textbf {\bibinfo {volume} {137}},\
  \bibinfo {pages} {074103} (\bibinfo {year} {2012})}\BibitemShut {NoStop}%
\bibitem [{\citenamefont {Rudzinski}\ \emph {et~al.}(2016)\citenamefont
  {Rudzinski}, \citenamefont {Kremer},\ and\ \citenamefont
  {Bereau}}]{rudzinski2016communication}%
  \BibitemOpen
  \bibfield  {author} {\bibinfo {author} {\bibfnamefont {J.~F.}\ \bibnamefont
  {Rudzinski}}, \bibinfo {author} {\bibfnamefont {K.}~\bibnamefont {Kremer}}, \
  and\ \bibinfo {author} {\bibfnamefont {T.}~\bibnamefont {Bereau}},\
  }\href@noop {} {\bibfield  {journal} {\bibinfo  {journal} {The Journal of
  Chemical Physics}\ }\textbf {\bibinfo {volume} {144}} (\bibinfo {year}
  {2016})}\BibitemShut {NoStop}%
\bibitem [{\citenamefont {Dixit}\ and\ \citenamefont {Dill}(2018)}]{CCMM}%
  \BibitemOpen
  \bibfield  {author} {\bibinfo {author} {\bibfnamefont {P.~D.}\ \bibnamefont
  {Dixit}}\ and\ \bibinfo {author} {\bibfnamefont {K.~A.}\ \bibnamefont
  {Dill}},\ }\href@noop {} {\bibfield  {journal} {\bibinfo  {journal} {Journal
  of chemical theory and computation}\ }\textbf {\bibinfo {volume} {14}},\
  \bibinfo {pages} {1111} (\bibinfo {year} {2018})}\BibitemShut {NoStop}%
\bibitem [{\citenamefont {Wan}\ \emph {et~al.}(2016)\citenamefont {Wan},
  \citenamefont {Zhou},\ and\ \citenamefont {Voelz}}]{MaxCalMutation}%
  \BibitemOpen
  \bibfield  {author} {\bibinfo {author} {\bibfnamefont {H.}~\bibnamefont
  {Wan}}, \bibinfo {author} {\bibfnamefont {G.}~\bibnamefont {Zhou}}, \ and\
  \bibinfo {author} {\bibfnamefont {V.~A.}\ \bibnamefont {Voelz}},\ }\href@noop
  {} {\bibfield  {journal} {\bibinfo  {journal} {Journal of chemical theory and
  computation}\ }\textbf {\bibinfo {volume} {12}},\ \bibinfo {pages} {5768}
  (\bibinfo {year} {2016})}\BibitemShut {NoStop}%
\bibitem [{\citenamefont {Dixit}\ \emph {et~al.}(2015)\citenamefont {Dixit},
  \citenamefont {Jain}, \citenamefont {Stock},\ and\ \citenamefont
  {Dill}}]{dixit2015inferring}%
  \BibitemOpen
  \bibfield  {author} {\bibinfo {author} {\bibfnamefont {P.~D.}\ \bibnamefont
  {Dixit}}, \bibinfo {author} {\bibfnamefont {A.}~\bibnamefont {Jain}},
  \bibinfo {author} {\bibfnamefont {G.}~\bibnamefont {Stock}}, \ and\ \bibinfo
  {author} {\bibfnamefont {K.~A.}\ \bibnamefont {Dill}},\ }\href@noop {}
  {\bibfield  {journal} {\bibinfo  {journal} {Journal of chemical theory and
  computation}\ }\textbf {\bibinfo {volume} {11}},\ \bibinfo {pages} {5464}
  (\bibinfo {year} {2015})}\BibitemShut {NoStop}%
\bibitem [{\citenamefont {Evans}\ \emph {et~al.}(2010)\citenamefont {Evans},
  \citenamefont {Searles},\ and\ \citenamefont
  {Williams}}]{evans2010probability}%
  \BibitemOpen
  \bibfield  {author} {\bibinfo {author} {\bibfnamefont {D.~J.}\ \bibnamefont
  {Evans}}, \bibinfo {author} {\bibfnamefont {D.~J.}\ \bibnamefont {Searles}},
  \ and\ \bibinfo {author} {\bibfnamefont {S.~R.}\ \bibnamefont {Williams}},\
  }\href@noop {} {\bibfield  {journal} {\bibinfo  {journal} {The Journal of
  chemical physics}\ }\textbf {\bibinfo {volume} {132}},\ \bibinfo {pages}
  {024501} (\bibinfo {year} {2010})}\BibitemShut {NoStop}%
\bibitem [{\citenamefont {Evans}\ \emph {et~al.}(2016)\citenamefont {Evans},
  \citenamefont {Williams}, \citenamefont {Searles},\ and\ \citenamefont
  {Rondoni}}]{evans2016relaxation}%
  \BibitemOpen
  \bibfield  {author} {\bibinfo {author} {\bibfnamefont {D.~J.}\ \bibnamefont
  {Evans}}, \bibinfo {author} {\bibfnamefont {S.~R.}\ \bibnamefont {Williams}},
  \bibinfo {author} {\bibfnamefont {D.~J.}\ \bibnamefont {Searles}}, \ and\
  \bibinfo {author} {\bibfnamefont {L.}~\bibnamefont {Rondoni}},\ }\href@noop
  {} {\bibfield  {journal} {\bibinfo  {journal} {arXiv preprint
  arXiv:1602.05808}\ } (\bibinfo {year} {2016})}\BibitemShut {NoStop}%
\bibitem [{\citenamefont {Crooks}(2011)}]{crooks2011thermodynamic}%
  \BibitemOpen
  \bibfield  {author} {\bibinfo {author} {\bibfnamefont {G.~E.}\ \bibnamefont
  {Crooks}},\ }\href@noop {} {\bibfield  {journal} {\bibinfo  {journal}
  {Journal of Statistical Mechanics: Theory and Experiment}\ }\textbf {\bibinfo
  {volume} {2011}},\ \bibinfo {pages} {P07008} (\bibinfo {year}
  {2011})}\BibitemShut {NoStop}%
\bibitem [{\citenamefont {Crooks}(1998)}]{CrooksMicRev}%
  \BibitemOpen
  \bibfield  {author} {\bibinfo {author} {\bibfnamefont {G.~E.}\ \bibnamefont
  {Crooks}},\ }\href@noop {} {\bibfield  {journal} {\bibinfo  {journal}
  {Journal of Statistical Physics}\ }\textbf {\bibinfo {volume} {90}},\
  \bibinfo {pages} {1481} (\bibinfo {year} {1998})}\BibitemShut {NoStop}%
\bibitem [{\citenamefont {Zhang}\ \emph {et~al.}(2012)\citenamefont {Zhang},
  \citenamefont {Qian},\ and\ \citenamefont {Qian}}]{zhang2012stochastic}%
  \BibitemOpen
  \bibfield  {author} {\bibinfo {author} {\bibfnamefont {X.-J.}\ \bibnamefont
  {Zhang}}, \bibinfo {author} {\bibfnamefont {H.}~\bibnamefont {Qian}}, \ and\
  \bibinfo {author} {\bibfnamefont {M.}~\bibnamefont {Qian}},\ }\href@noop {}
  {\bibfield  {journal} {\bibinfo  {journal} {Physics Reports}\ }\textbf
  {\bibinfo {volume} {510}},\ \bibinfo {pages} {1} (\bibinfo {year}
  {2012})}\BibitemShut {NoStop}%
\bibitem [{\citenamefont {Maes}\ \emph {et~al.}(2007)\citenamefont {Maes},
  \citenamefont {Neto{\v{c}}n{\`y}},\ and\ \citenamefont
  {Wynants}}]{maes2007and}%
  \BibitemOpen
  \bibfield  {author} {\bibinfo {author} {\bibfnamefont {C.}~\bibnamefont
  {Maes}}, \bibinfo {author} {\bibfnamefont {K.}~\bibnamefont
  {Neto{\v{c}}n{\`y}}}, \ and\ \bibinfo {author} {\bibfnamefont
  {B.}~\bibnamefont {Wynants}},\ }\href@noop {} {\bibfield  {journal} {\bibinfo
   {journal} {arXiv preprint arXiv:0709.4327}\ } (\bibinfo {year}
  {2007})}\BibitemShut {NoStop}%
\bibitem [{\citenamefont {Bereau}\ and\ \citenamefont
  {Swendsen}(2009)}]{bereau2009optimized}%
  \BibitemOpen
  \bibfield  {author} {\bibinfo {author} {\bibfnamefont {T.}~\bibnamefont
  {Bereau}}\ and\ \bibinfo {author} {\bibfnamefont {R.~H.}\ \bibnamefont
  {Swendsen}},\ }\href@noop {} {\bibfield  {journal} {\bibinfo  {journal}
  {Journal of Computational Physics}\ }\textbf {\bibinfo {volume} {228}},\
  \bibinfo {pages} {6119} (\bibinfo {year} {2009})}\BibitemShut {NoStop}%
\bibitem [{\citenamefont {Donati}\ \emph {et~al.}(2017)\citenamefont {Donati},
  \citenamefont {Hartmann},\ and\ \citenamefont {Keller}}]{donati2017girsanov}%
  \BibitemOpen
  \bibfield  {author} {\bibinfo {author} {\bibfnamefont {L.}~\bibnamefont
  {Donati}}, \bibinfo {author} {\bibfnamefont {C.}~\bibnamefont {Hartmann}}, \
  and\ \bibinfo {author} {\bibfnamefont {B.~G.}\ \bibnamefont {Keller}},\
  }\href@noop {} {\bibfield  {journal} {\bibinfo  {journal} {The Journal of
  chemical physics}\ }\textbf {\bibinfo {volume} {146}},\ \bibinfo {pages}
  {244112} (\bibinfo {year} {2017})}\BibitemShut {NoStop}%
\bibitem [{\citenamefont {Ma}\ \emph {et~al.}(2017)\citenamefont {Ma},
  \citenamefont {Su}, \citenamefont {Lai},\ and\ \citenamefont
  {Tong}}]{ma2017colloidal}%
  \BibitemOpen
  \bibfield  {author} {\bibinfo {author} {\bibfnamefont {X.-g.}\ \bibnamefont
  {Ma}}, \bibinfo {author} {\bibfnamefont {Y.}~\bibnamefont {Su}}, \bibinfo
  {author} {\bibfnamefont {P.-Y.}\ \bibnamefont {Lai}}, \ and\ \bibinfo
  {author} {\bibfnamefont {P.}~\bibnamefont {Tong}},\ }\href@noop {} {\bibfield
   {journal} {\bibinfo  {journal} {Physical Review E}\ }\textbf {\bibinfo
  {volume} {96}},\ \bibinfo {pages} {012601} (\bibinfo {year}
  {2017})}\BibitemShut {NoStop}%
\bibitem [{\citenamefont {Seifert}(2005)}]{seifert2005entropy}%
  \BibitemOpen
  \bibfield  {author} {\bibinfo {author} {\bibfnamefont {U.}~\bibnamefont
  {Seifert}},\ }\href@noop {} {\bibfield  {journal} {\bibinfo  {journal}
  {Physical review letters}\ }\textbf {\bibinfo {volume} {95}},\ \bibinfo
  {pages} {040602} (\bibinfo {year} {2005})}\BibitemShut {NoStop}%
\bibitem [{\citenamefont {Bolhuis}\ \emph {et~al.}(2002)\citenamefont
  {Bolhuis}, \citenamefont {Chandler}, \citenamefont {Dellago},\ and\
  \citenamefont {Geissler}}]{bolhuis2002transition}%
  \BibitemOpen
  \bibfield  {author} {\bibinfo {author} {\bibfnamefont {P.~G.}\ \bibnamefont
  {Bolhuis}}, \bibinfo {author} {\bibfnamefont {D.}~\bibnamefont {Chandler}},
  \bibinfo {author} {\bibfnamefont {C.}~\bibnamefont {Dellago}}, \ and\
  \bibinfo {author} {\bibfnamefont {P.~L.}\ \bibnamefont {Geissler}},\
  }\href@noop {} {\bibfield  {journal} {\bibinfo  {journal} {Annual review of
  physical chemistry}\ }\textbf {\bibinfo {volume} {53}},\ \bibinfo {pages}
  {291} (\bibinfo {year} {2002})}\BibitemShut {NoStop}%
\end{thebibliography}%

\end{document}